\documentclass[iop]{emulateapj}

\shorttitle{Visualisation of multi-mission astronomical data with ESASky}
\shortauthors{Baines et al.}

\begin{document}


\title{Visualisation of multi-mission astronomical data with ESASky}


\author{Deborah Baines\altaffilmark{1}, Fabrizio Giordano\altaffilmark{1}, Elena Racero\altaffilmark{1}, Jes\'{u}s Salgado\altaffilmark{1}, Bel\'{e}n L\'{o}pez Mart\'{i}\altaffilmark{1,2}, Bruno Mer\'{i}n\altaffilmark{1}, Mar\'{i}a-Henar Sarmiento\altaffilmark{1}, Ra\'{u}l Guti\'{e}rrez\altaffilmark{1}, I\~{n}aki Ortiz de Landaluce\altaffilmark{1}, Ignacio Le\'{o}n\altaffilmark{1}, Pilar de Teodoro\altaffilmark{1}, Juan Gonz\'{a}lez\altaffilmark{1}, Sara Nieto\altaffilmark{1}, Juan Carlos Segovia\altaffilmark{1}, Andy Pollock\altaffilmark{1,3}, Michael Rosa\altaffilmark{1}, Christophe Arviset\altaffilmark{1}, Daniel Lennon\altaffilmark{1}, William O'Mullane\altaffilmark{1} and Guido de Marchi\altaffilmark{4}.} 


\altaffiltext{1}{European Space Astronomy Centre (ESAC), European Space Agency (ESA), 28691 Villanueva de la Ca\~{n}ada, Spain}
\altaffiltext{2}{Saint Louis University - Madrid campus, Avenida del Valle 34, 28003 Madrid, Spain}
\altaffiltext{3}{Department of Physics and Astronomy, University of Sheffield, Hicks Building, Hounsfield Road, Sheffield S3 7RH, UK}
\altaffiltext{4}{European Space Research and Technology Centre, Keplerlaan 1, 2201 AZ Noordwijk, The Netherlands}

\begin{abstract}
ESASky is a science-driven discovery portal to explore the multi-wavelength sky and visualise and access multiple astronomical archive holdings. The tool is a web application that requires no prior knowledge of any of the missions involved and gives users world-wide simplified access to the highest-level science data products from multiple astronomical space-based astronomy missions plus a number of ESA source catalogues. The first public release of ESASky features interfaces for the visualisation of the sky in multiple wavelengths, the visualisation of query results summaries, and the visualisation of observations and catalogue sources for single and multiple targets. This paper describes these features within ESASky, developed to address use cases from the scientific community. The decisions regarding the visualisation of large amounts of data and the technologies used were made in order to maximise the responsiveness of the application and to keep the tool as useful and intuitive as possible.\newline


\end{abstract}

\section{Introduction}
Many astronomy data centres and archiving facilities are currently addressing the need from the scientific community to provide intuitive, fast and visual access to their data holdings, with the main goal of maximising the exploitation of scientific data. This need comes from a number of issues, such as the yearly exponential increase in astronomical data; the need to understand astronomical objects across the full electromagnetic spectrum and/or across the time domain; the need for easy access to science-ready data (or the highest-level data products) from as many astronomical observatories as possible; and to understand which observatories, if any, have observed in a particular region of the sky. 

The ESAC Science Data Centre (ESDC), located at the European Space Astronomy Centre (ESAC) near Madrid, Spain, hosts the archives from ESA space science missions (astronomy, planetary science and heliophysics) and provides services and tools to access and retrieve ESA mission data. With this experience, and to fulfill the requests from the scientific community, the ESDC have developed ESASky (http://sky.esa.int), a web-based data discovery portal for visualising and accessing science-ready astronomical data in collective astronomical archives (\citealp{BM:2015}; \citealp{BM:2016}). The tool was developed in collaboration and with support from many ESA mission science and technical experts, and from the Centre de Donn\'{e}es astronomiques de Strasbourg (CDS). ESASky provides full access to the entire sky as observed by ESA (and other) missions, and provides a simple way to search and download multi-wavelength data, going beyond the individual archives boundaries. 

Version 1 of ESASky was released in May 2016 and can be found at the following URL: \newline

\texttt{http://sky.esa.int} \newline

This version addresses three main use cases: (i) the exploration of the multi-wavelength skies; (ii) the search and retrieval of data for single targets; and (iii) the search and retrieval of data for multiple target lists. In addition to the three main use cases, a fourth is also addressed: the visualisation of the sky coverage from all missions. This first version contains imaging data sets and catalogues only (future releases will enable retrieval of spectra and will have special time domain exploration features).

In this paper, we describe the visualisation features within version 1 of ESASky, developed to address the use cases; the technologies used to develop these features; and the decisions made regarding the visualisation of large amounts of data, with the aim of keeping the tool as useful and intuitive as possible. Section \ref{sec:sky} describes the visualisation and exploration of the sky and section \ref{sec:histograms} describes how the results are summarised in histograms. Sections \ref{sec:obs} and \ref{sec:cats} describe the visualisation of the observations and catalogues respectively, and finally, the summary is given in section \ref{sec:sum}.  

\pagebreak

\section{Visualisation of the sky}\label{sec:sky}
Before developing ESASky, a strong user requirement was to make the tool web-based. Google Web Toolkit\footnote{http://www.gwtproject.org} (GWT; a development toolkit for building and optimising complex browser-based applications) was chosen as the development technology for ESASky, the decision being based on a feasability study by ESDC into web-based technologies and on previous experience developing archives with GWT. 

The second decision to be made was how to fulfill the use case to explore multi-wavelength skies. For this, we chose Aladin Lite \citep{TB:2014} developed by CDS, a lightweight version of the Aladin Sky Atlas \citep{FB:2000}, running in the browser and geared towards simple visualisation of the sky. To integrate Aladin Lite into the ESASky source code, we developed a GWT wrapper using the JavaScript Native Interface (JSNI) methods to communicate directly with the JavaScript of Aladin Lite. 

Aladin Lite displays progressive multi-resolution, all-sky projections of full mission data sets using HEALPix projections \citep{KG:2005} called Hierarchical Progressive Surveys (HiPS), also developed by CDS \citep{PF:2015a}. Hundreds of HiPS from major astronomical surveys and observatories have been created by CDS and other institutes, and they can be visualised within Aladin Lite (see http://aladin.u-strasbg.fr/hips/list for a comprehensive list). As the user zooms into the sky the spatial resolution of the HiPS changes until the lowest level is reached, which is normally the resolution of the instrument. To keep ESASky simple and easy to use, we chose to show a subset of the available HiPS (some of the most well-known surveys and missions) and we created our own HiPS for the ESA astronomy missions. This was achieved using the Aladin/hipsgen\footnote{http://aladin.u-strasbg.fr/hips/} code, a package used to create progressive HiPS maps from a collection of calibrated FITS files. It should be noted that the ESA HiPS are intended for visualisation only and are not science-ready products and should therefore not be used to perform scientific data analysis. ESASky instead provides easy access to the highest-level data products from each mission (see section \ref{sec:obs}). ESA astronomy mission HiPS have been created for the Akari, EXOSAT, Herschel, Hubble Space Telescope (HST), INTEGRAL, Infrared Space Observatory (ISO), Planck and XMM-Newton missions and are available through ESASky (see the ``skies/HiPS'' section of the online ESASky help for details, accessible from http://sky.esa.int). 

\begin{figure*}
\epsscale{1.17}
\plotone{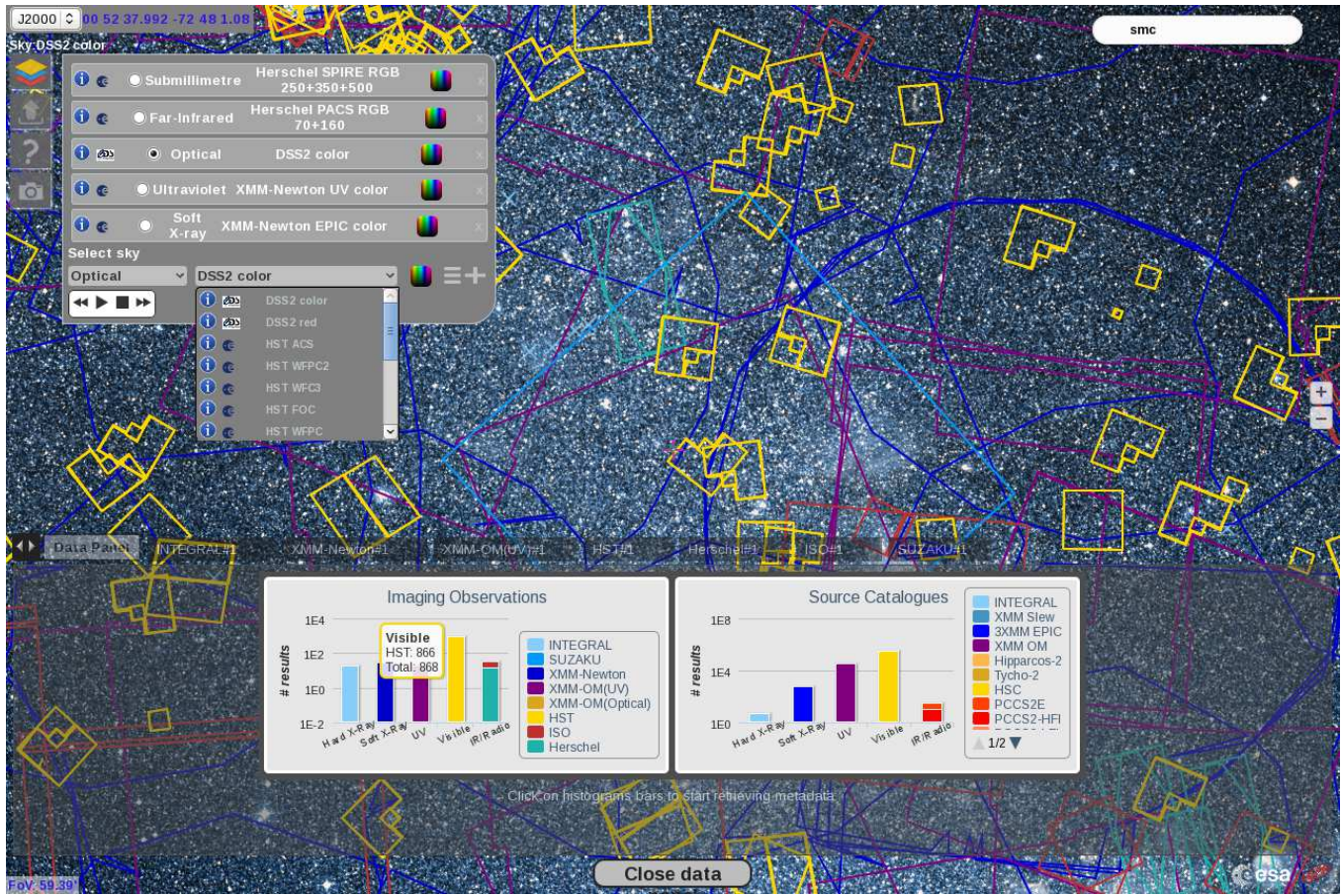}
\caption{ESASky visualising the centre of the Small Magellanic Cloud as seen in the optical with the Digitized Sky Survey (DSS2). The skies menu is open in the top left of the tool and includes a number of stacked skies. Results histograms are shown towards the bottom of the tool with an example of hovering over a bar to display the number of available observations (in this case for the HST). Observation footprints are shown for Herschel (cyan), HST (yellow), ISO (red), XMM-Newton (X-ray: dark blue; ultraviolet: purple), SUZAKU (blue), and pointings for INTEGRAL (light blue crosses). \label{fig:ESASkyHistograms}}
\end{figure*}

The HiPS are available to visualise through ESASky via the skies menu and are displayed in the Equatorial system (J2000) by default (the user can also choose the Galactic coordinate system). The skies menu was chosen to be mission agnostic, and groups the skies (HiPS) by wavelength regions, from $\gamma$-ray to radio. The skies can be stacked above the menu and played through using video style buttons to aid the user exploration experience. The user can also modify the colour of the skies using the colour map options: greyscale, rainbow, EOSB, Planck and reverse. Figure \ref{fig:ESASkyHistograms} shows the skies menu open in ESASky with a number of stacked skies. Transparency of one or more skies will be available in the near future.

\section{Visualisation of the results summary}\label{sec:histograms}

To fulfill the use case to search and retrieve data for single targets, the results of a search first need to be visualised. ESASky contains a search field that accepts coordinates (equatorial or Galactic) or an object name that can be resolved by SIMBAD.\footnote{http://simbad.u-strasbg.fr/simbad/} The search region displayed in ESASky is chosen by taking the size of the object from SIMBAD. All the available archives are then queried in this region (within the corner coordinates of the display).

The results of this query are summarised in histograms, one for the observations and one for the catalogues, to give the user a general overview of all results available. The results shown are the highest-level science observations only. The histograms are ordered by wavelength (similar to the skies menu), assuming that the average user does not know the specific details of every mission and catalogue available, but is interested in wavelength regions (time domain searches will be available in a future version). The histograms are interactive, and by hovering over the histogram bars the user can see a summary showing the number of observations and sources for each mission or catalogue respectively. By clicking on the mission names in the legends, the histogram bars can be hidden to better visualise others; for example, those with small numbers of observations or sources. The histograms are automatically refreshed and re-scaled when one does this or when one moves to a different region of the sky. These histograms are produced with the JavaScript Highcharts\footnote{http://www.highcharts.com} library. Figure \ref{fig:ESASkyHistograms} shows an example of the results histograms (centred on the Small Magellanic Cloud). 

To maximise the responsiveness of the tool, a threshold is placed at a field of view (FOV) of three degrees or larger, after which the number of observations and catalogue sources are obtained from pre-computed tables (rather than directly on the fly from the individual archives). These tables were created by dividing the sky (represented as a sphere) into equal area HiPS pixels, with each pixel containing a pre-computed number of observations and sources per mission and catalogue respectively. An approximate count is then calculated from the pre-computed tables and displayed in the results histograms. These tables are updated regularly for missions still in operation (such as HST and XMM-Newton). 


\begin{figure*}
\epsscale{1.17}
\plotone{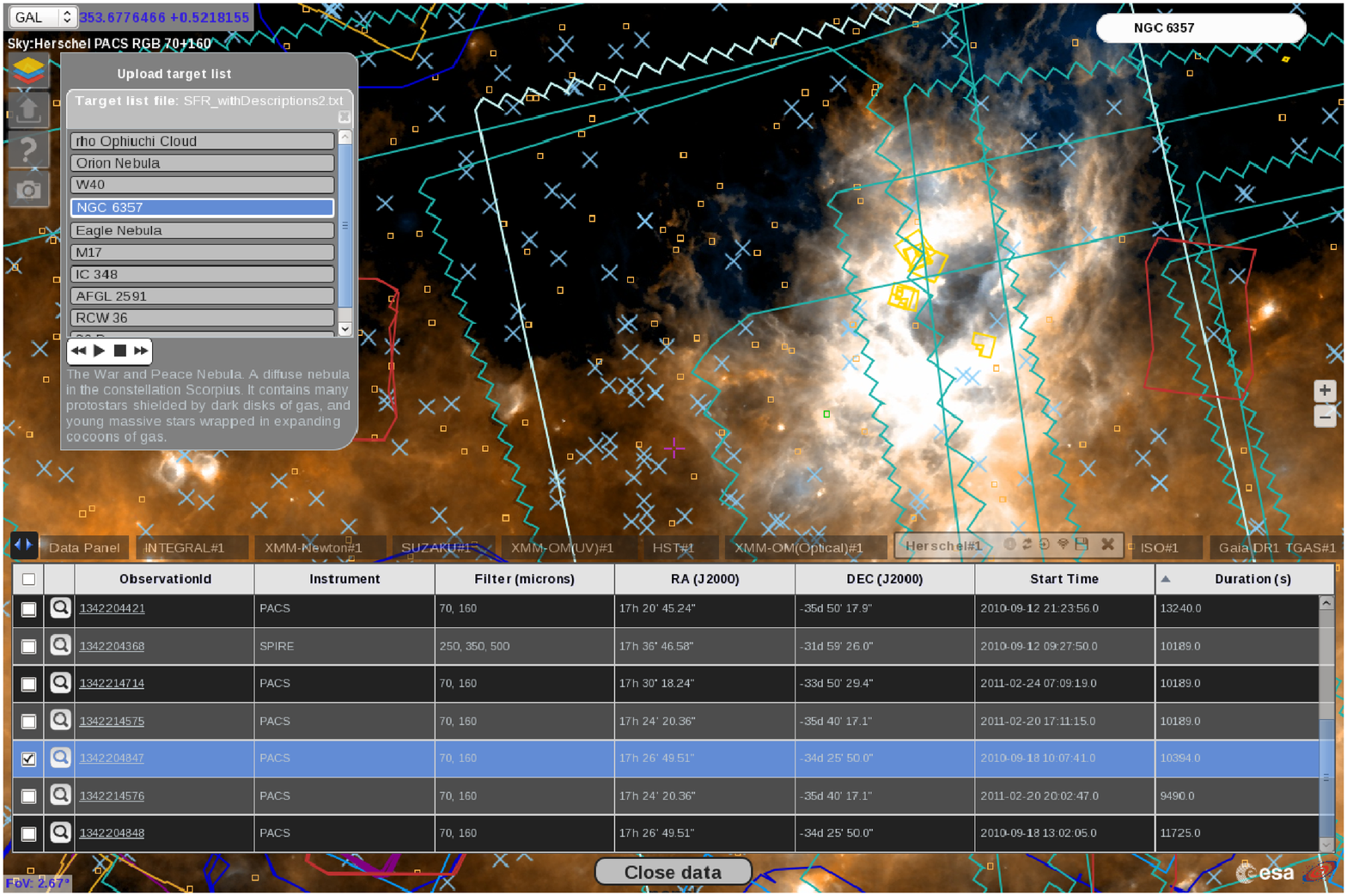}
\caption{ESASky visualising NGC 6357 as seen in the far-infrared with the Herschel PACS RGB HiPS. The multi-target menu is open in the top left of the tool and includes an uploaded target list file and text. The results table for Herschel is shown towards the bottom of the tool, selecting one row (in blue) highlights the observation footprint on the sky. Observation footprints are shown for Herschel (cyan), HST (yellow), ISO (red), XMM-Newton (X-ray: dark blue; ultraviolet: purple; optical: orange), SUZAKU (blue), and pointings for INTEGRAL (light blue crosses). Catalogue sources are shown from the Gaia TGAS catalogue (\citealp{DM:2015}; orange squares). \label{fig:ESASkyObs}}
\end{figure*}

\section{Visualisation of the observations}\label{sec:obs}

In version 1 of ESASky, imaging observations are available from the ESA Herschel, HST, INTEGRAL, ISO and XMM-Newton missions and from the JAXA SUZAKU mission (see http://sky.esa.int help for details). Details on the observations are loaded in ESASky via the summary histograms by clicking on the histogram bars. The archives are queried in the region displayed, a results table is provided per mission, and the footprints of each observation are displayed on the sky (or in the case of INTEGRAL, the central pointing positions are displayed; see below for details). 

The footprints for each observation are generated by the missions and obtained from the individual mission archives. In the case where no footprints were available (ISO and XMM-Newton), we generated the footprints using the ST-ECF Footprint Finder\footnote{http://hla.stsci.edu/Footprintfinder/FootprintFinder.html} code, with support from the instrument experts. In ESASky, each mission has a different coloured footprint to easily distinguish one mission from another in the sky. 

The results of a query are summarised in a table per mission, where each row is a separate observation. Metadata such as the observation ID number, the instrument, the coordinates, amongst others, are displayed in the table, as chosen by the mission experts. The observational data themselves (and also the catalogue data) remain in their corresponding archives, and ESASky provides the user with access to download the data products from these archives. For all missions, direct links are available to the image previews. In the case of HST and XMM-Newton, a link is provided (via the observation ID) that opens the corresponding archive results page for each specific observation. Users can then explore further in the archives to obtain more detailed information for each observation. A similar link is provided for SUZAKU, and we plan on extending this to other archives in future releases. 

The results table can also be downloaded, either as a comma-separated value (CSV) file or a VOTable \citep{FO:2009} file and it can be sent via the Simple Application Messaging Protocol (SAMP; \citealp{MT:2015}; \citealp{MT:2012}) to SAMP enabled tools (such as TOPCAT; \citealp{MT:2005}). In addition to providing the column metadata, the product url, postcard url and the footprint coordinates are also provided for each observation. Finally, for nearly all the missions, the imaging data can also be sent via SAMP to SAMP enabled image visualisation tools, such as DS9 \citep{WJ:2003} and Aladin \citep{FB:2000}. 

In the case of the ESA INTEGRAL $\gamma$-ray mission, a different approach was needed to display the observations from the IBIS (Imager on Board the INTEGRAL Satellite; \citealp{PU:2003}) instrument. The total FOV of the instrument (down to zero response) is 29.1$^{\circ}$ x 29.4$^{\circ}$. It is not helpful to the user to visualise the whole footprint on the sky for two reasons: (i) because the instrument sensitivity decreases towards the edge of the FOV and (ii) because the footprints cover such a large area of the sky, and in some regions there are many pointings close to one another, it visually becomes confusing what regions have been observed. Instead, we display the central coordinates of each pointing as a (light blue) cross. Figures \ref{fig:ESASkyHistograms} and \ref{fig:ESASkyObs} include examples of INTEGRAL pointings visualised on the sky. 

The number of footprints displayed on the sky is limited to a few thousand (this number is optimised per mission), again in order to maximise the performance of the tool. Above this threshold, a multi-order coverage map (MOC; \citealp{PF:2015b}) for each mission is displayed. A MOC is a HEALPix pixelization of the coverage maps of complete surveys or missions and is a simple way to map regions of the sky into hierarchically grouped predefined cells. For all the ESA missions in ESASky, we created MOCs simultaneously with the HiPS maps using the HIPSGEN code within Aladin (see \citealt{BLM:2016} for more details on the MOCs created for ESASky). Visualising the MOCs of every mission provides a fast and easy way to compare mission coverages against one another over the entire sky, and addresses the use case to visualise the sky coverage from all missions. 

To fulfill the use case to search and retrieve data for multiple target lists, ESASky allows the user to upload a text file containing a list of targets, either names resolved by SIMBAD or a list of coordinates. Once loaded, the target list appears in the tool and is interactive: targets can be selected and the sky will reload, centred on the target; or one can play through the list of targets and view the summary of results; or multiple results tabs can be opened for the same mission, or catalogue, and different targets. In addition, text and comments can be added to the target list file, which are then displayed in ESASky below the target list. This is a useful feature for users to keep notes on specific objects of interest, as well as being useful for education purposes: users can be taken on a tour of astronomical objects in multi-wavelength skies. Figure \ref{fig:ESASkyObs} shows an example target list and text displayed in ESASky.


\section{Visualisation of catalogues}\label{sec:cats}

Version 1 of ESASky provides the visualisation of 11 ESA mission catalogues, from the Gaia, Hipparcos, HST, INTEGRAL, Planck and XMM-Newton missions (see http://sky.esa.int help for details). The catalogues vary in sizes, from just a few thousand rows, to tens of millions of rows (in the case of the Hubble Source Catalog, HSC; \citealp{BW:2016}). Two main issues occur when visualising such large catalogues: (i) the speed of the tool decreases; and (ii) the user can be overloaded with visual information and individual sources can't be resolved. To prevent this we apply two filters, firstly, a warning is given to refine the search by using a smaller FOV when there are more than 50,000 sources in the FOV. Secondly, in the case of more than 2000 sources in the FOV, ESASky orders them by magnitude or flux and shows the brightest 2000. An exception to this is for the HSC, where the 2000 displayed sources are ordered by the most observed sources (the NumImages value). A higher NumImage value provides a higher confidence in the magnitude values and can eliminate artifacts present in the catalogue.  

As with the observations, the catalogues are loaded via the interactive histograms. A choice was made by the ESA mission specialists on the most important columns to display within the results table per catalogue. For the majority of catalogues, these columns are source name, right ascension, declination, flux or magnitude, errors and source flags, but other columns are also shown depending on the type of catalogue (for example, proper motions and parallax for Hipparcos and TGAS). ESASky provides interaction between the results table and the sky, and vice versa, i.e. clicking on a row in the results table will highlight the source on the sky and clicking on the source on the sky will highlight the results table row and will bring up a summary box. The summary box provides the source name, coordinates, which catalogue (results tab name) it belongs to, and links to SIMBAD, the NASA/IPAC Extragalactic Database\footnote{https://ned.ipac.caltech.edu} (NED), and the VizieR photometry viewer\footnote{http://vizier.u-strasbg.fr/vizier/sed/}, where searches are performed centred on the source coordinates. As with the observations, the catalogues can be downloaded or sent to a VO application via SAMP. In the case of the XMM-Newton 3XMM-DR6 catalogue \citep{SR:2016}, one can obtain further catalogue information via a link in the results table which opens the XMM-Newton Science Archive results for each source. We plan to provide this functionality for all catalogues in the near future.


\section{Summary}\label{sec:sum}

As a response to our archive users, we have developed ESASky, a web-based science-driven discovery portal to visualise and access multiple astronomical archive holdings. The first public release of ESASky features interfaces for the visualisation of the sky in multiple wavelengths, the visualisation of query results summaries, and the visualisation of observations and catalogue sources for single and multiple targets. This version addresses use cases from the scientific community, such as the exploration of multi-wavelength skies; searching and retrieving data for single targets; searching and retrieving data for multiple target lists; and visualising the sky coverage from all missions. 

In addition to addressing the use cases, the choices that were made on the visualisation technologies and techniques used were for four main reasons: (i) to maximise the responsiveness of the application; (ii) to keep the tool simple and intuitive; (iii) to not overload the user with too much visual information; and (iv) to enhance the user experience.  

ESASky (and the underlying ESA astronomy archives) makes use of many International Virtual Observatory Alliance\footnote{http://ivoa.net} (IVOA) protocols and notes, such as HiPS, MOC, TAP, STC-S, SAMP, VOTable etc. These protocols play an essential part in the tool, enhance the user experience, and have helped to increase the speed at which we have been able to develop ESASky. 

The next major release of ESASky will include the retrieval of spectroscopic data, links to more of the mission catalogues and improvements in some of the visualisation functionalities. After this, future functionalities will include spectral visualisation and special time domain exploration features.


\vspace{3mm}


Acknowledgements: We acknowledge the excellent support from the expert science and technical staff at ESAC and CDS. In particular, we acknowledge the following people: Mark Allen, Bruno Altieri, Ruben \'{A}lvarez, Guillaume Belanger, Thomas Boch, Tam\'{a}s Budav\'{a}ri, Javier Castellanos, Xavier Dupac, Daniel Durand, Ken Ebisawa, Matthias Ehle, Pilar Esquej, Pierre Fernique, Pedro Garc\'{i}a Lario, Krzysztof G\'{o}rski, Jonas Haase, John Hoar, Peter Kretschmar, Erik Kuulkers, Ren\'{e} Laureijs, Nora Loiseau, Marcos L\'{o}pez-Caniego, Alejandro Lorca, Anthony Marston, Antonella Nota, G\"{o}ran Pilbratt, Roberto Prieto, Pedro Rodr\'{i}guez, Miguel S\'{a}nchez-Portal, Maria Santos-Lleo, Norbert Schartel, Jan Tauber, Ivan Valtchanov and Eva Verdugo. ESASky makes use of Aladin Lite, developed at CDS, Strasbourg Observatory, France \citep{TB:2014}. This work also benefited from experience gained from projects supported by the Mission Operation Division of ESA's Directorate of Science.


\end{document}